\documentclass[twocolumn]{aastex631}

\begin{document}

\title{Toward Microarcsecond Astrometry for the Innermost Wobbling Jet of the BL Lacertae Object OJ 287}

\correspondingauthor{Xiaopeng Cheng}
\email{xcheng0808@gmail.com}
\correspondingauthor{Jun Yang}
\email{jun.yang@chalmers.se}

\author[0000-0003-4407-9868]{Xiaopeng Cheng}
\affiliation{Korea Astronomy and Space Science Institute, 776 Daedeok-daero, Yuseong-gu, Daejeon 34055, Korea}

\author[0000-0002-2322-5232]{Jun Yang}
\affiliation{Department of Earth and Space Sciences, Chalmers University of Technology, Onsala Space Observatory, SE-43992 Onsala, Sweden}

\author[0000-0002-4417-1659]{Guangyao Zhao}
\affiliation{Instituto de Astrof\'{\i}sica de Andaluc\'{\i}a---CSIC, Glorieta de la Astronom\'{\i}a s/n, E-18008~Granada, Spain}

\author[0000-0002-4148-8378]{Bong Won Sohn}
\affiliation{Korea Astronomy and Space Science Institute, 776 Daedeok-daero, Yuseong-gu, Daejeon 34055, Korea}
\affiliation{University of Science and Technology, Gajeong-ro 217, Yuseong-gu, Daejeon 34113, Republic of Korea}

\author[0000-0001-7003-8643]{Jung Taehyun}
\affiliation{Korea Astronomy and Space Science Institute, 776 Daedeok-daero, Yuseong-gu, Daejeon 34055, Korea}
\affiliation{University of Science and Technology, Gajeong-ro 217, Yuseong-gu, Daejeon 34113, Republic of Korea}

\author[0000-0002-9093-6296]{Xiaofeng Li}
\affiliation{Department of Astronomy, Guangzhou University, Guangzhou 510006, PR China}
\affiliation{Great Bay Center, National Astronomical Data Center, Guangzhou, Guangdong 510006, China}
\affiliation{Astronomy Science and Technology Research Laboratory of Department of Education of Guangdong Province, Guangzhou 510006, China}








\begin{abstract}

The BL Lacertae object OJ~287 is a very unusual quasar producing a wobbling radio jet and some double-peaked optical outbursts with a possible period of about 12~yr for more than one century. This variability is widely explained by models of binary supermassive black hole (SMBH) or precessing jet/disk from a single SMBH. To enable an independent and nearly bias-free investigation on these possible scenarios, we explored the feasibility of extremely high-precision differential astrometry on its innermost restless jet at mm-wavelengths. Through re-visiting some existing radio surveys and very long baseline interferometry (VLBI) data at frequencies from 1.4 to 15.4~GHz and performing new Very Long Baseline Array (VLBA) observations at 43.2~GHz, we find that the radio source J0854$+$1959, 7.1 arcmin apart from OJ~287 and no clearly-seen optical and infrared counterparts, could provide a nearly ideal reference point to track the complicated jet activity of OJ~287. The source J0854$+$1959 has a stable GHz-peaked radio spectrum and shows a jet structure consisting of two discrete, mas-scale-compact and steep-spectrum components and showing no proper motion over about 8 yr. The stable VLBI structure can be interpreted by an episodic, optically thin and one-sided jet. With respect to its 4.1-mJy peak feature at 43.2~GHz, we have achieved an astrometric precision at the state-of-art level, about 10~$\mu$as. These results indicate that future VLBI astrometry on OJ~287 could allow us to accurately locate its jet apex and activity boundary, align its restless jet structure over decades without significant systematic bias, and probe various astrophysical scenarios.

\end{abstract}

\keywords{: Very long baseline interferometry (1769) --- Supermassive black holes (1663) --- Radio jets (1347) --- Radio continuum emission (1340)}


\section{Introduction} \label{sec:intro}

OJ~287 is a well-known low-synchrotron peaked (LSP) BL Lacertae object at redshift z = 0.306 \citep{Sitko1985, Stickel1989}. It is one of the most promising active galactic nuclei (AGN) that shows repeated double-peaked outburst features in the optical regime with $\sim$ 12 yr intervals \citep{Sillanpaa1988}. To explain this double-peaked quasi-periodic oscillation (QPO) feature, \citet{Sillanpaa1988} proposed a binary supermassive black hole (SMBH) system at the center. Based on the binary SMBH model, several scenarios were proposed and investigated such as a mildly or highly precessing secondary SMBH and a precessing accretion disk or jet associated with primary SMBH \citep{Sillanpaa1988, Lehto1996, Katz1997, Valtonen2012, Britzen2018}. The detailed model parameters such as the mass of the primary SMBH are still in updating \citep{Liu2002, Valtonen2012, Dey2018, Komossa202302, Komossa202304}.
However, other models that do not necessarily require the presence of a secondary SMBH are also explored to explain the optical and radio light curves \citep[e.g.,][]{Britzen2018, Li2018, Butuzova2020}.

The early radio observations of OJ~287 by \citet{Gabuzda1989} with Very Long Baseline Interferometry (VLBI) technique revealed that its jet components have a superluminal motion. After that, there are more complex radial motions found in some jet components \citep[e.g. stationary and decelerating components, ][]{Vicente1996}. The long-term VLBI observations of OJ~287 show that its jet has a complex transverse motion \citep[e.g.][]{Tateyama2004, Agudo2012, Tateyama2013, Cohen2017, Britzen2018}, e.g. erratic wobbling with $\sim$150$\degr$ change of the position angle (PA) between 2004 and 2006 \citep{Agudo2012}. The wobbling behavior is very prominent in the inner sub-mas jet region and provides significant support for precessing jet scenarios \citep[e.g.][]{Britzen2018}. Because of the rapid structure evolution \citep[e.g. proper motion 0.2--0.4~mas~yr$^{-1}$,][]{Britzen2018} and some significant flux density variability \citep[e.g.][]{Komossa202304}, there is no stationary reference point available to accurately align these sub-mas-resolution VLBI images observed at more than 200 epochs and over about three decades. 
With advanced calibration technique of the tropospheric delay \citep{Reid2014}, the astrometry precision from some existing phase-referencing observations at 22~GHz has reached $\sim$10~$\mu$as per epoch \citep[e.g.][]{Zhang2013}. Currently, it is still quite challenging to perform high-precision VLBI differential astrometry at $>$22~GHz \citep[e.g.][]{Guirado2000, Sokolov2008, Abellan2018} because of very short phase coherence time ($<$60~s for typical weather condition) and missing nearby ($<30\arcmin$) compact and stable sources \citep[e.g. a recent review by][]{Rioja2020}. To constrain precessing jet models for OJ~287 via monitoring the jet PA in particular in the sub-mas-scale region, it is necessary to accurately locate the innermost jet origin that might be undetectable because of strong absorption or variability and have significant offsets from the map peak features.

Given that OJ~287 is a very bright source with a flux density of 1--10~Jy at mm wavelengths, a nearby faint source can be used to provide an external reference point to track its erratic wobbling motion. To enable extremely high-precision astrometry on the innermost restless jet of OJ~287, we have thoroughly investigated the goodness of the candidate reference source J0854$+$1959. It has an angular separation of 7.1~arcmin to OJ~287. To date, there is no firm identification of its optical and infrared counterparts.
At radio bands, J0854$+$1959 is unresolved in the 1.4 GHz Very Large Array (VLA) Faint Images of the Radio Sky at Twenty-Centimeters \citep[FIRST;][]{Becker1995} image and Very Long Baseline Array (VLBA) image \citep[][]{Deller2014}.

This Letter is organized in the following sequence. Section \ref{sec:obs} describes our VLBI observations and data reduction and the data analysis for the available archival VLBI data. Section \ref{result} presents these multi-frequency and multi-epoch VLBI imaging results and the broad-band radio spectrum of J0854$+$1959. We interpret its jet structure and discuss its goodness for future state-of-the-are astrometry on the wobbling jet of OJ~287 and possible direct scientific implication in Section \ref{discuss}. We give our conclusions in Section~\ref{summary}.

\section{OBSERVATIONS AND DATA REDUCTION} \label{sec:obs}

\subsection{VLBA observation at 43.2~GHz}\label{2.1}

The VLBA observation (project code: BY166, PI: Jun Yang) of J0854$+$1959 and OJ 287 was conducted on 09 February 2022 at 43.2~GHz with a recording rate of 4096~Mbps (dual polarization, 4 $\times$ 128~MHz sub-bands per polarization, 2-bit quantization). The duration time was about 3~h. Except for the Hancock station, all the VLBA stations participated in the observation. The raw VLBA data were correlated using the DiFX software correlator \citep{Deller2007, Deller2011} at Socorro, NM, with an averaging time of 2~s and a frequency resolution of 0.5~MHz.

Because of the weakness of J0854$+$1959, we used the inverse phase-referencing observations \citep{Beasley1995} with a cycle time of 40~s: 15~s for two gaps, 5~s for OJ~287, and 20~s for J0854$+$1959. The on-source time for J0854$+$1959 and OJ~287 was about 86 and 22 min. Thanks to the small angular separation ($\sim$7 arcmin), we allocate 15s for switching between sources in each cycle. The correlation position for OJ~287 is RA = 08$^{\rm h}$54$^{\rm m}$48$^{\rm s}$.8749270; DEC. = 20$^{\rm \degr}$06$^{\rm '}$30$^{\rm ''}$.640851 ($\rm \sigma _{ra}$ = 0.03 mas, $\rm \sigma _{dec}$ = 0.03 mas) provided by the radio fundamental catalogue 2016A \citep{2020A&A...644A.159C}. The phase center of J0854+1959 is RA = 08$^{\rm h}$54$^{\rm m}$54$^{\rm s}$.9156550; DEC. = 19$^{\rm \degr}$59$^{\rm '}$34$^{\rm ''}$.856100 provided by the radio fundamental catalogue 2020B \citep{Petrov2021}. During the post-data reduction, OJ~287 was used as the fringe finder and bandpass calibrator.

The data were calibrated using the US National Radio Astronomy Observatory (NRAO) Astronomical Image Processing System \citep[AIPS;][]{Greisen2003} in a standard procedure. A-prior amplitude calibration was carried out using the system temperatures and antenna gains measured at each station during the observation. The dispersive delays caused by the ionosphere were corrected from a map of total electron content provided by the Global Positioning System satellite observations and the earth orientation parameters were corrected using the measurements for the U.S. Naval Observatory database. Phase errors due to the antenna parallactic angle variations were removed. The instrumental single-band delays and phase offsets were corrected using 2 minutes of observational data from OJ 287. After inspecting the data and flagging, global fringe fitting was performed on OJ~287 with a 5-s solution interval and a point-source model by averaging over all the sub-bands. We then applied the phase solutions from OJ~287 to J0854$+$1959 by linear interpolation. 

\begin{figure}
    \centering
    \includegraphics[width=0.52\textwidth]{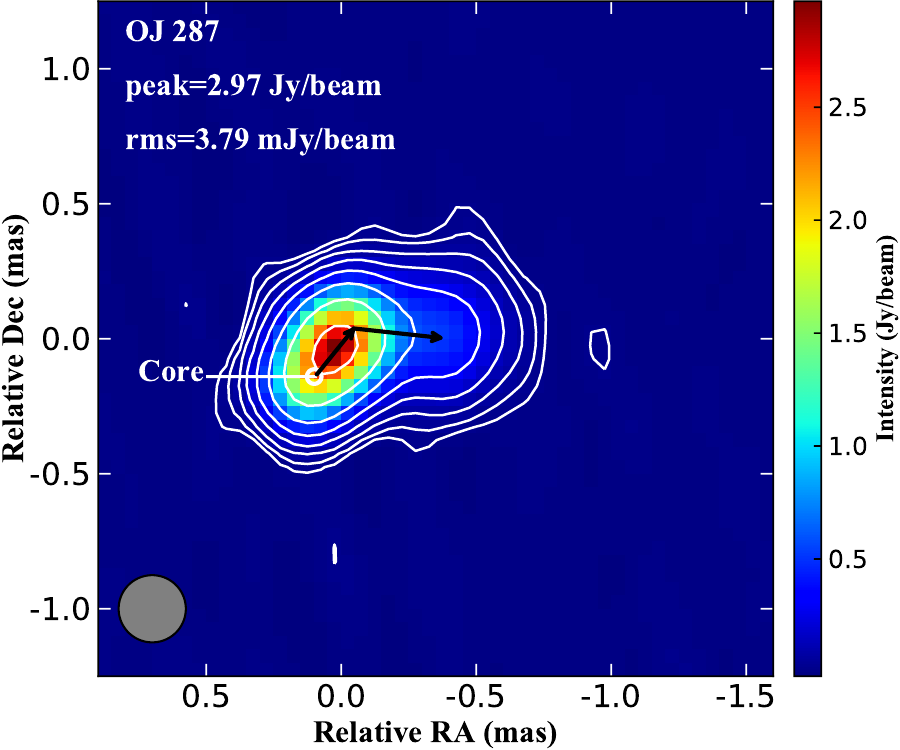}
    \includegraphics[width=0.46\textwidth]{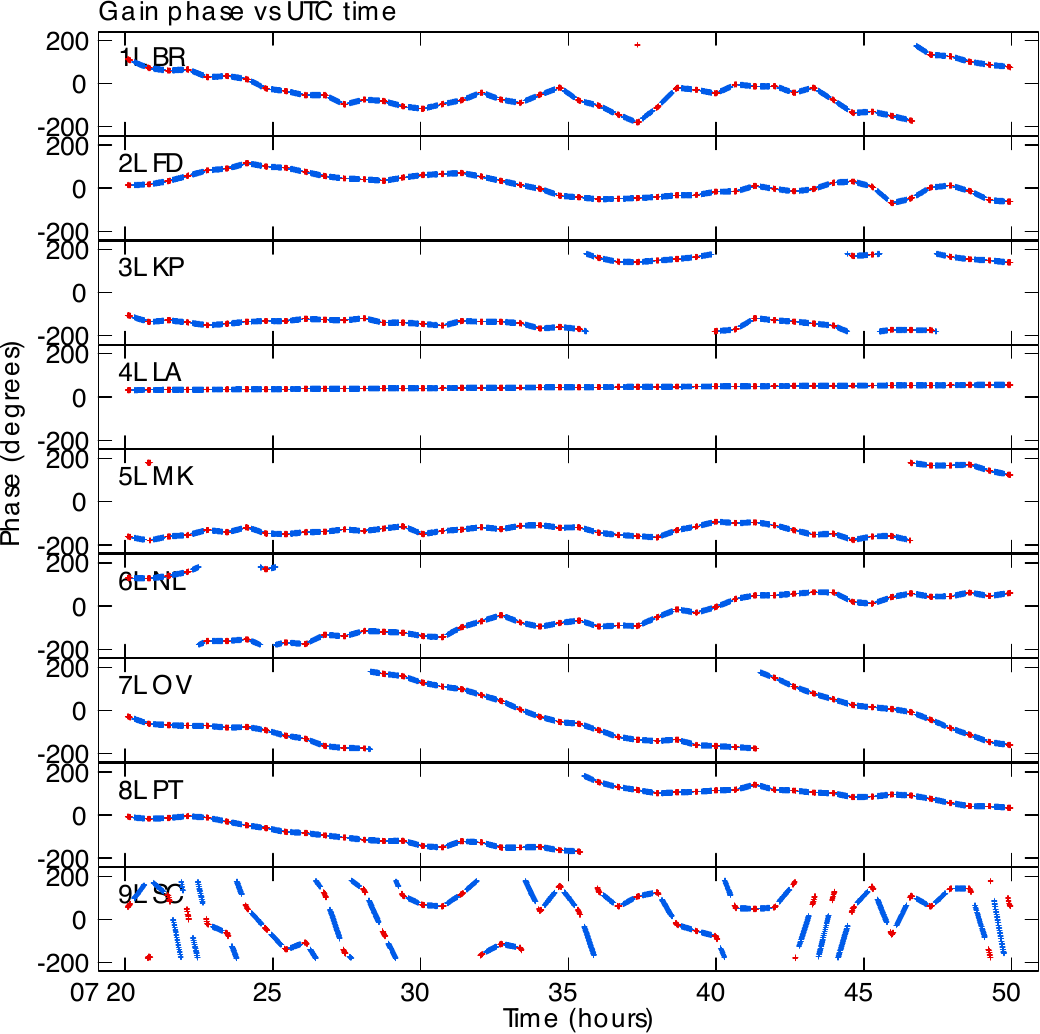}
    \caption{The core-jet structure of OJ 287 and the accurate phase connection between OJ~287 and J0854$+$1959 in our phase-referencing VLBA observations at 43 GHz. In the left panel, the lowest contour is at $\sim$3 $\sigma$ level. The image is made with uniform weighting and a circular Gaussian beam, 0.159 $\times$ 0.159 mas$\rm ^{2}$. The black arrow represents the evolution of the jet PA with distance. In the right panel, red symbols are the phase solutions of OJ 287; blue symbols are the phase interpolation results from OJ~287 to J0854$+$1959. \label{fig:phas}}
\end{figure}

We first imaged OJ 287, see Fig. \ref{fig:phas}, in Difmap \citep{Shepherd1997}. The source shows a compact one-sided core-jet structure along the southeast to northwest direction with a PA of $-$43$\degr$ in the inner $\sim$ 0.4 mas. The jet underwent a sharp jet bend of 50$\degr$ at $\sim$ 0.4 mas distance from the core. Our image is consistent with the recent 43 and 86 GHz VLBI images \citep{Britzen2018,2022A&A...658L..10L,2022ApJ...932...72Z}. We iteratively ran model fitting with point source and self-calibration to the visibility data. Both the solutions of the amplitude self-calibration and the phase dependent on the source structure were obtained and applied to both sources in AIPS. Fig. \ref{fig:phas} displays the final phase solutions of the first half hour. The calibrator OJ~287 and the target J0854$+$1959 are plotted as red and blue symbols, respectively. The rapid phase variation at SC was because the station is on an island and had wet weather. The high signal-to-noise ratio of the solutions and the mild temporal evolution of the phases guarantee the successful application of inverse phase-referencing. The final visibility data of J0854+1959 were split out and imaged in Difmap. Two weak circular Gaussian components ($\sim$ 10.0 $\sigma$ and $\sim$ 5.5 $\sigma$ for S1 and S0, respectively) were fitted in Difmap using the MODELFIT program. The final image was created in natural weighting. The integrated flux density and precise position of each component are listed in Table \ref{astro}.

\subsection{Archival VLBI data}

We also searched for the archival data of J0854$+$1959 in some VLBI databases and found 4 experiments at 1.4-15 GHz. Table~\ref{model} lists some basic information about these VLBI experiments. The calibrated visibility data for the VLBA projects BD161AS (PI: A.T. Deller) and BR235F (PI: L. Petrov) were downloaded from the mJy Imaging VLBA Exploration at 20~cm \citep[mJIVE-20, ][]{Deller2014} site \footnote{http://safe.nrao.edu/vlba/mjivs/products.html} and the Astrogeo database\footnote{VLBA calibrator survey database is maintained by Leonid Petrov: http://astrogeo.org/}, respectively. The correlation data of the European VLBI Network (EVN) project EP081A (PI: R.W. Porcas) and the VLBA project BP167B (PI: R.W.~Porcas) were downloaded from the EVN data archive\footnote{http://archive.jive.nl/scripts/arch.php?exp=EP081A} and the National Radio Astronomy Observatory Archive\footnote{https://data.nrao.edu/portal/}. The data of EP081A at 5~GHz and BP167B at 2.3, 5.0, and 8.4 GHz were analyzed with AIPS and Difmap following the standard VLBI data reduction procedures. The correlated data of BP167B at 15.4 GHz were reduced with AIPS and Difmap in the standard phase-referencing VLBI data reduction procedures (the same procedure described in Section \ref{2.1}). 
All the data were imaged and modeled in Difmap to quantitatively describe the emission structure. Further information regarding the observations can be found in Table \ref{model}.

\section{RESULTS}\label{result}

\subsection{Radio Morphology of J0854+1959}

\begin{figure*}
\centering
 \includegraphics[height=4.5cm]{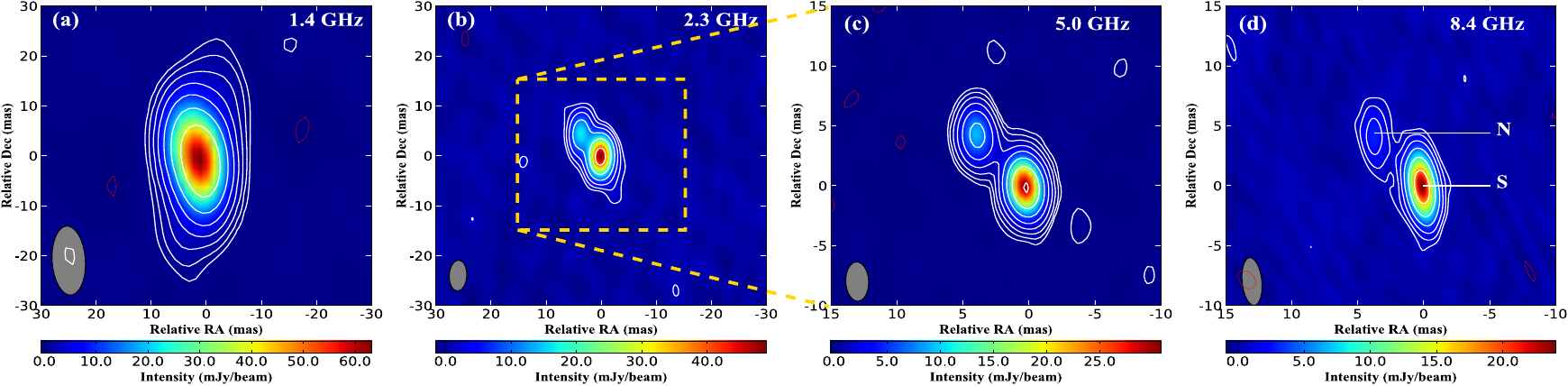}
\\
\caption{The compact double jet structure of J0854$+$1959 was observed with the VLBA at frequencies from 1.4 to 8.4 GHz on 2012 March 7 and September 14. All the images are made with natural weighting. The lowest contour in all the images is at 3 $\sigma$ level. The inner yellow box of the 2.3-GHz image indicates the size of the two higher-frequency images. (a) The VLBA image at 1.4 GHz. The restoring beam has a full width at half-maximum (FWHM) of 12.14 $\times$ 4.64 mas$\rm ^{2}$ at a PA of 0.79$\degr$. The counters are 0.52 $\times$ ($-$1, 1, 2, ..., 64) mJy beam$^{-1}$. (b) The VLBA image at 2.3 GHz. The beam FWHM is 6.18 $\times$ 3.18 mas$\rm ^{2}$ at a PA of $-$2.48$\degr$. The counters are 1.24 $\times$ ($-$1, 1, 2, 4, ..., 32) mJy beam$^{-1}$. (c) The VLBA image at 5.0 GHz. The beam FWHM is 3.27 $\times$ 1.74 mas$\rm ^{2}$ at a PA of 2.05$\degr$. The counters are 0.28 $\times$ ($-$1, 1, 2, 4, ..., 64) mJy beam$^{-1}$. (d) The VLBA image at 8.4 GHz. The beam FWHM is 2.66 $\times$ 1.12 mas$\rm ^{2}$ at a PA of 16.80$\degr$. The counters are 0.54 $\times$ ($-$1, 1, 2, 4, ..., 32) mJy beam$^{-1}$. 
} \label{fig:VLBI_low}
\end{figure*}

\begin{figure*}
\centering
 \includegraphics[width=0.31\textwidth]{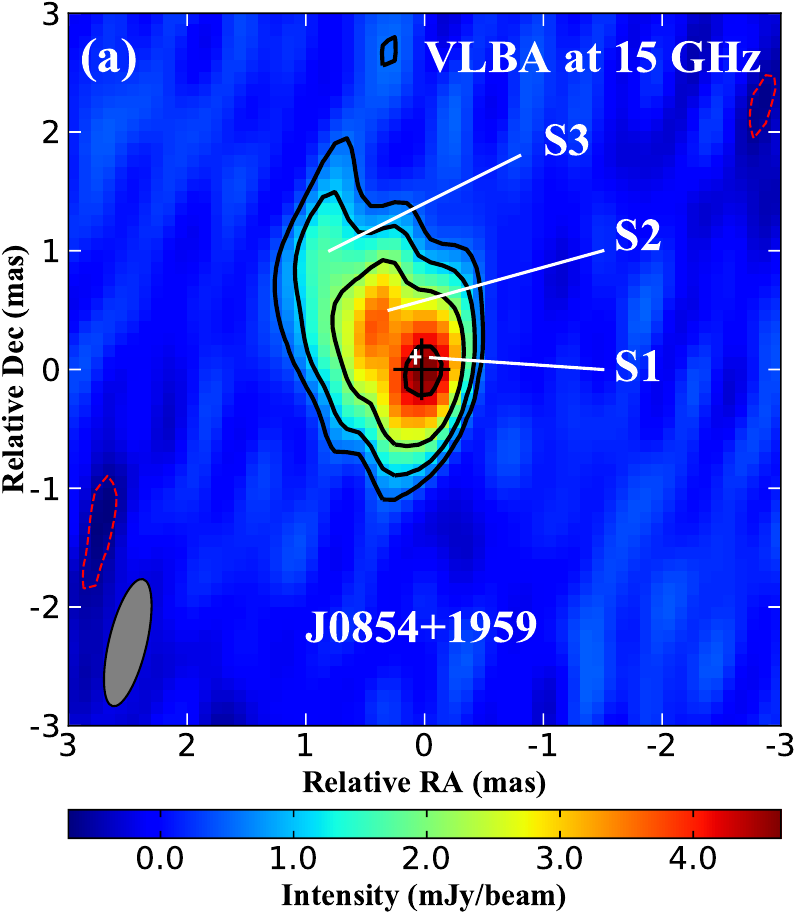}
 \includegraphics[width=0.32\textwidth]{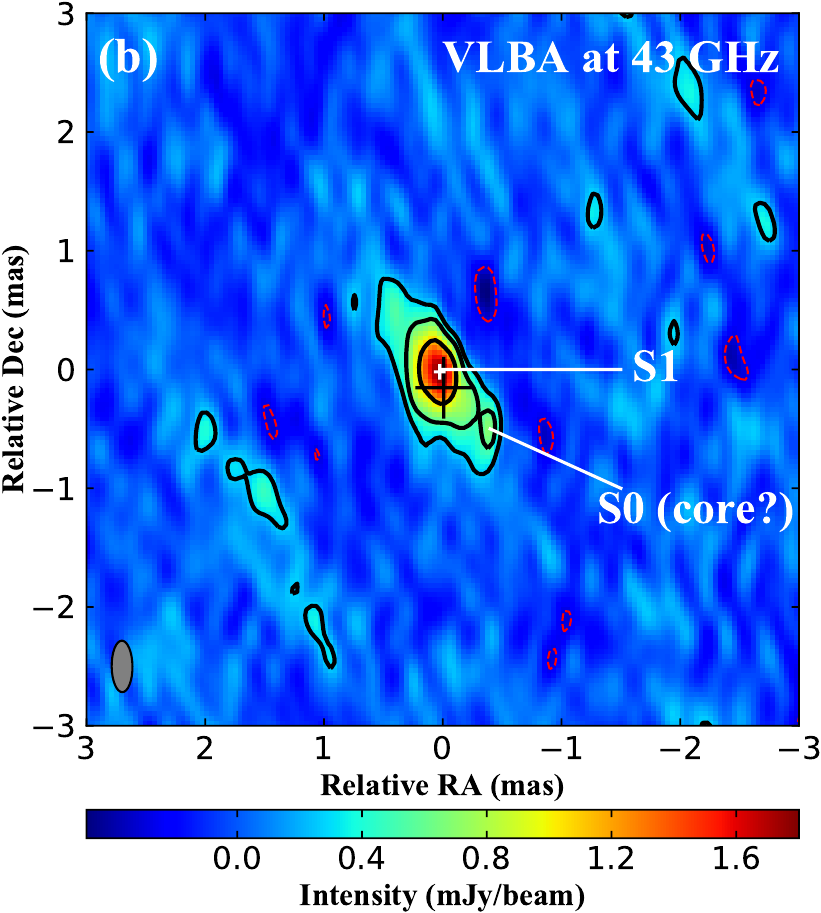}
 \includegraphics[width=0.34\textwidth]{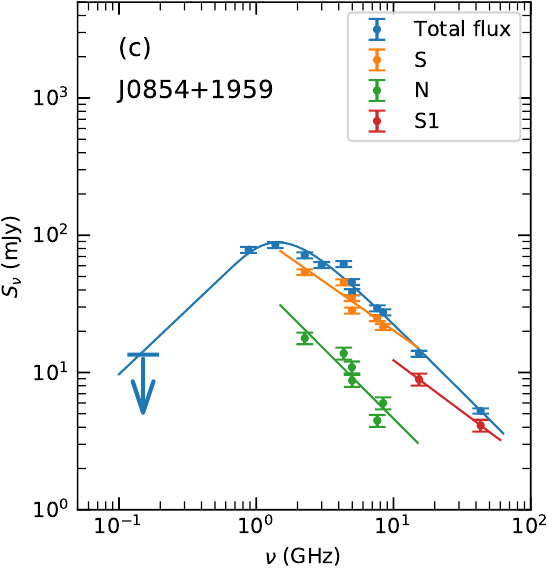}
\\
\caption{The inner structure of the southern component and the non-simultaneous spectra of J0854$+$1959. In the two VLBA images observed over one decade and aligned with the radio core of OJ~287 as the reference, the centroids of the steady component S1 are marked as a white cross at 43 GHz and a black cross at 15 GHz. Their sizes represent the 5 $\sigma$ error bars. (a) The 15 GHz image is made with natural weighting. The beam FWHM is 1.10 $\times$ 0.32 mas$\rm ^{2}$ at a PA of $-$13.45$\degr$. The lowest contour in the image is at 3 $\sigma$ level. The counters are 0.51 $\times$ ($-$1, 1, 2, 4, 8) mJy beam$^{-1}$. (b) The 43 GHz image is made with natural weighting. The beam FWHM is 0.43 $\times$ 0.18 mas$\rm ^{2}$ at a PA of 0.58$\degr$. The lowest contour in the image is at $\sim$2.5 $\sigma$ level. The counters are 0.275 $\times$ ($-$1, 1, 2, 4) mJy beam$^{-1}$. (c) The blue curve shows the best-fitting results of total flux using equation \ref{eq1}; the green, orange, and red lines show the best-fitting results of the components N, S, and S1. The data collection and the models of the broad-band radio spectra are described in section \ref{3.2}.
} \label{fig:VLBI_high}
\end{figure*}

Figure \ref{fig:VLBI_low} shows the naturally weighted total intensity VLBI images at 1.4--8.4 GHz. The compact double structure in the northeast-southwest direction looks like a Compact Symmetric Object (CSO) \citep{2012ApJS..198....5A}, but the southern component S is much brighter and more compact than the northern component N. J0854$+$1959 was observed with the VLBA simultaneously at $\sim$ 5 and 8 GHz on 14 September 2012 and 04 November 2020, respectively. According to the results of the model fitting of components N and S, listed in Table \ref{model}, the spectral index ($\rm S_{\nu} \propto \nu^{\alpha}$, where $S_{\nu}$ is specific flux density, $\nu$ is the observing frequency) for N and S are estimated. The mean values of N and S are $-$1.00$\pm$0.22 and $-$0.7$\pm$0.14, indicating two steep-spectrum components. At 15 GHz, component N is not detected, suggesting a steep-spectrum and extended structure, and component S is resolved into 3 components, S1--S3, in the northeast direction (Fig. \ref{fig:VLBI_high}(a)), consistent with the previous observations at low frequencies. Our 43 GHz image in Fig. \ref{fig:VLBI_high}(b) reveals a fine inner jet structure.  Two components are detected within 0.5 mas: S1 at the same position within 5$\sigma$ as the component at 15 GHz, and a new component S0 in the upstream of the jet direction.

\subsection{Radio spectra} \label{3.2}

J0854$+$1959 was not detected at 150 MHz in the TIFR GMRT Sky Survey \citep[TGSS;][]{Intema2017}, which gives an upper limit of 13.5 mJy (detection threshold of 5$\sigma$). The total flux is about 78.4 mJy at 887 MHz with an unresolved core in the Rapid Australian Square Kilometre Array Pathfinder (ASKAP) Continuum Survey \citep[RACS;][]{McConnell2020}. The 1.4 GHz VLA  image shows that the source is still compact with a total flux of 85.1 mJy. On mas scales, the source is resolved into two components (S and N). The southern component S is resolved into two or three components at 15 and 43 GHz. The model flux measurements of each component at 1.4--43.2 GHz are listed in Table \ref{model} and \ref{astro}.

Since J0854+1959 has a wide spectral coverage from 0.07--43.2 GHz and a convex spectrum that peaks at gigahertz frequencies in Fig.~\ref{fig:VLBI_high}(c), it has a similarity to a GHz-peaked spectrum (GPS) source \citep[e.g.][]{ODea1998} with a steep spectral slope at the high frequencies. The error bars correspond to the total uncertainty in the measured flux densities, as a result of the 10\% amplitude calibration uncertainty. According to the synchrotron self-absorption model for the spherical homogeneous plasma, we fit the radio spectrum of the source by adopting the following function given by \citet{2017ApJ...836..174C}
\begin{equation}
\tiny
S_{\rm \nu}(\nu) = \frac{S_{\rm p}}{1-\exp(-1)} 
\left( \frac{\nu}{\nu_{\rm p}} \right)^{\alpha_{\rm thick}} \left \{ 1 -\exp \left[ - \left( \frac{\nu}{\nu_{\rm p}} \right)^{\alpha_{\rm thin} - \alpha_{\rm thick}} \right] \right \},  
\label{eq1} 
\end{equation}
where $\alpha_{\rm thin}$ and $\alpha_{\rm thick}$ are the spectral indices in the optically thin and thick region, respectively. $S_{\rm p}$ is the flux density at the spectral turnover frequency $\nu_{\rm p}$, $\nu$ is the observing frequency. When $\alpha_{\rm thick}=2.5$, this mode represents a spherical homogeneous synchrotron self-absorption source. The least-square fitting gives $S_{\rm p}=87.63\pm5.86$ mJy,  $\nu_{\rm t}=1.64\pm0.20$ GHz, $\alpha_{\rm thin}=-1.00\pm0.05$, $\alpha_{\rm thick}=0.95\pm0.14$. 

We also fitted the data of components N and S with the standard non-thermal power-law model of the form
\begin{equation}
S_{\rm \nu}(\nu) = S_{\rm 0}\nu^{\rm \alpha}
\label{eq2} 
\end{equation}
Where S$_{\rm 0}$ is the characterizes of the amplitude of the synchrotron spectrum, $\alpha$ is the synchrotron spectral index, and S$_{\rm \nu}$ is the flux density at frequency $\nu$, respectively. The fitting gives S$_{\rm 0} = 46.37\pm16.81$, $\alpha = -1.00\pm0.22$ for component N and S$_{\rm 0} = 101.90\pm22.77$, $\alpha = -0.70\pm0.14$ for component S, respectively.

\section{Discussion}\label{discuss}

\subsection{The nature of the radio continuum structure}

From the radio images and spectra, J0854+1959 could be identified as a one-sided jet GPS source or a CSO. None of these two components are identified as the flat-spectrum radio core because of the optical thin spectra ($\alpha_{\rm S}$ = $-$0.7 and $\alpha_{\rm N}$ = $-$1.0). The new component S1 between 15 and 43 GHz also has a steep spectrum with $\alpha$ = $-$0.75. The absence of a flat-spectrum core makes the morphology identification less certain. We fitted the Gaussian component position of component N with respect to the bright component S at each epoch and only found a hint for the positional difference, $\sim$0.05 mas over 8 years. We used a simple non-acceleration, two-dimensional vector motion on the 5 GHz data \citep{2021MNRAS.506.1609C}. The separation of N and S is at a rate of 4.5$\pm$24.0 $\mu$as/yr, indicating no significant proper motion found within the error. Because of the steep spectrum ($-$0.75) and almost no position changes (0.12 mas) between 15 and 43 GHz at about 10 years, S1 is identified as a stationary jet component with position changing less than 12 $\mu$as/yr. A new component S0 is marginally detected with an SNR of about 5.5$\sigma$ at the outer edge of component S. New VLBI observations are required to clarify its physical nature. If we confirm it in our future observations, it suggests S0 is the active nucleus of J0854+1959. Following the above information, we invoke a core–jet interpretation of the source structure, similar to other core-jet sources \citep[e.g., J0017+5312:][]{2012ApJS..198....5A}. In this source, the reference component S is actually a bright stationary component with very low proper motion (4.5$\pm$24.0 $\mu$as/yr). The separation speed reflects the relative motion of the two main jet knots and not the real absolute motion of the source. The possible reason for the none-detection of the radio core is that the core is obscured by the bright inner jet component and dimmed due to either synchrotron self-absorption or free–free absorption at low frequencies \citep{2010MNRAS.402...87A,2014ApJ...780..178M}.

To constrain the jet viewing angle and intrinsic jet speed, we identify the components S and N as the approaching jet components and give an upper limit of the none-detection of the receding jet components based on the image sensitivity. Based on the assumption that all the components are emitting the same radio luminosity isotropically in their respective rest frames, the flux density ratio $R_{\rm flux}$ between approaching and receding jets is \citep[e.g.,][]{2012rjag.book.....B}
\begin{equation}
R_{\rm flux}=\frac{S_{\rm a}}{S_{\rm r}} = \left(  \frac{1 + \beta\cos\theta_{\rm v}}{1-\beta\cos\theta_{\rm v}} \right) ^{3-\alpha},
\label{eq4}
\end{equation}
where $S_{\rm a}$ and $S_{\rm r}$ are the flux densities of the approaching and receding components respectively,  $\beta$ is the intrinsic jet speed in the unit of the light speed c, $\theta_{\rm v}$ is the viewing angle between the jet axis and the line of sight, $\alpha=-1.04\pm0.05$ is the spectral index of the jet. If we assume the total flux of the approaching jet is 84.86 mJy at 1.4 GHz and the upper limit of the total flux in the receding jet is 0.54 mJy ($\sim 3\sigma$), we can drive a lower limit of the ratio of 157 and a constrain of $\beta\cos\theta_{\rm v}$ = $0.56\, c$. This gives $\beta >$ $0.55\, c$, $\theta_{\rm v} <$ 56$\degr$.The deriving jet speed suggests that J0854+1959 shows a one-sided moderate relativistic jet in mas scales. 

Most CSOs show symmetric morphology, edge-brightening, and steep spectra of double components, while the Central radio cores are too weak to be detected \citep[e.g.,][]{2003PASA...20...69P,2012ApJS..198....5A,2020MNRAS.494.1744Y,2021A&ARv..29....3O}.
The separation velocities of the hot spots are relatively low.
A possibility is that some CSOs are core-jet structures, the two observed components are bright knots in a single-sided jet, while the core, which is not detected in the present observations, lies at one end of the jet, such as J0854+1959 in this paper.

\subsection{Toward extremely high-precision astrometry on the innermost wobbling jet of OJ287}

Because J0854+1959 has a very stable and mas-scale compact jet component, J0854+1959 can provide an almost ideal external reference point to track the wobbling jet of OJ 287. The extremely closeness ($\sim$7 arcmin) of the pair of sources can help to significantly reduce antenna slewing time and accurately mitigate systematic phase errors mainly resulting from the unstable troposphere in the differential astrometry \citep[e.g.][]{Reid2014}. This would allow VLBI phase-referencing observations at $\leq$43~GHz to reach their thermal noise limits, which only depend on the image resolution and signal-to-noise ratio (SNR). If more sensitive stations, e.g. the Green Bank, are added, the current astrometry precision of $\sim$10 $\mu$as per epoch can be boosted by a factor of two or more. Depending on the target precision, the advanced calibration of tropospheric delay \citep[e.g.][]{Reid2014} may be included.

The jet of OJ~287 has an extremely complex motion \citep[e.g.][]{Agudo2012, Britzen2018} and shows large radio flux density variability \citep{Komossa202302, Komossa202304}. The existence of a compact and stable reference point in J0854$+$1959 would help to answer some key questions for the inner sub-mas-scale wobbling jet. The significant PA change and the collimated inner jet structure could allow us to trace its jet apex from just two-epoch VLBI observations. The geometric jet apex represents the jet nozzle and might be optically thick. 
Compared with the centroid of the apparent radio core, the jet apex is free from any frequency-dependent opacity effect, i.e. core shift \citep[e.g.][]{Plavin2019}, and flux density variability. Long-term multi-epoch monitoring observations would allow us to reveal the evolution of the inner jet PA without significant bias, and find the boundary of the erratic jet activity. There were some tentative detections of a counter jet in OJ~287 at certain epochs in the previous VLBI observations \citep[e.g.][]{Agudo2012}. The accurate determination of the jet apex could help us to firmly confirm the existence of a counter jet in future VLBI observations. The solid detection of a counter jet would enable us to tightly constrain the jet parameters. Because of frequently irregular outbursts, there might be new ejecta moving out and significantly changing the structure of the apparent radio core. These potential small structure changes could be also uncovered at the early phase by future extremely high-precision phase-referencing imaging observations. Moreover, J0854$+$1959 can enable the accurate determination of the small core shift \citep[$\rm \sim 30 \mu as$ between 8.1 and 15.4 GHz,][]{Pushkarev2012} of OJ~287. This would allow us to study the magnetic field in the innermost jet.  

In the future, the long-baseline mode of the next-generation VLA (ngVLA\footnote{\url{https://ngvla.nrao.edu/}}) could allow us to reach an image sensitivity of $\sim$1~$\mu$Jy\,beam$^{-1}$ at 30--50 GHz and thus achieve an astrometry precision of $\leq$1~$\mu$as per epoch. Such precision might also provide more direct astrometric constraints on the potential orbital motion of the primary SMBH. In literature, there also some similar efforts in other sources, e.g. 3C~66B \citep[][]{Sudou2003, Zhao2011}, BL~Lac \citep{Sokolov2008}, B0402$+$379 \citep{Bansal2017} and NGC~4472 \citep{Wrobel2022}. Compared with these projects, the VLBI astrometry with respect to J0854$+$1959 would be free from the core jitters from the reference source. 

\section{Summary}\label{summary}

To search for a suitable reference source for OJ 287 and directly measure the core shift and absolute precessing motion in the inner jet region, we searched the VLBI database and found a GPS source J0854+1959 with an angular separation of about 7 arcmin to OJ 287. J0854+1959 shows a moderate relativistic one-sided jet structure at low frequencies in the southwest to northeast direction. The previous VLBA phase-reference observation at 15 GHz shows that the north component was resolved and the southern component was resolved into 3 components (S1-S3). We carried out a new VLBA phase-reference observation of J0854+1959 and OJ 287 at 43 GHz. J0854+1959 was first successfully detected and imaged at 43 GHz. We obtained an astrometric accuracy $\sim$10 $\mu$as.  The 15 and 43 GHz phase-referencing observations results show that S1 is a steep spectrum and ultra-stable component over 10 yr, suggesting a stationary jet component. This suggests that S1 is an ideal reference point to perform mm-VLBI high-precision astrometry on OJ 287. A new component S0 is detected upstream of the jet. Future observations will confirm the nature of S0.

\begin{acknowledgments}
X.-P. Cheng and B.-W. Sohn were supported by Brain Pool Program through the National Research Foundation of Korea (NRF) funded by the Ministry of Science and ICT (2019H1D3A1A01102564). 
GYZ acknowledges financial support from the Severo Ochoa grant CEX2021-001131-S funded by MCIN/AEI/ 10.13039/501100011033 and the M2FINDERS project funded by the European Research Council (ERC) under the European Union’s Horizon 2020 research and innovation programme (grant agreement No 101018682). 
X.-F. Li was supported by the National Natural Science Foundation of China (No.12203014).
The EVN is a joint facility of independent European, African, Asian, and North American radio astronomy institutes. Scientific results from data presented in this publication are derived from the following EVN project code: EP081. The VLBA is an instrument of the National Radio Astronomy Observatory, a facility of the National Science Foundation operated under cooperative agreement by Associated Universities. 
\end{acknowledgments}

%

\vspace{5mm}
\facilities{VLBA, EVN.}


\software{AIPS \citep{Greisen2003}, DiFX \citep{Deller2007}, DIFMAP \citep{Shepherd1997}}

\bibliography{J0854}{}
\bibliographystyle{aasjournal}


\begin{deluxetable}{ccccccccc}
\centering
\tablecolumns{9}
\tabletypesize{\small}
\tablewidth{0pt}
\tablecaption{\label{astro} VLBA 15- and 43-GHz astrometry results of J0854+1959}
\tablehead{
\colhead{Freq.}  & \colhead{Component} & \colhead{S$_{\rm peak}$} & \colhead{S$_{\rm tot}$} & \colhead{Right Ascension} & \colhead{$\sigma_{\rm ra}$} & \colhead{Declination} & \colhead{$\sigma_{\rm dec}$} & \colhead{$\theta_{\rm size}$} \\
\colhead {(GHz)} & \colhead{(mas)}   & \colhead{(mJy beam$^{-1}$)}  & \colhead{(mJy)}   & \colhead{(J2000)}   & \colhead{(mas)}  & \colhead{(J2000)}     & \colhead{(mas)} & \colhead{(mas)}     \\
\colhead{(1)}    & \colhead{(2)}  & \colhead{(3)}  & \colhead{(4)}   & \colhead{(5)}   & \colhead{(6)}   & \colhead{(7)} & \colhead{(8)} & \colhead{(9)}}
\startdata
15.4 & S1 & 4.61$\pm$0.11 & 8.92$\pm$0.45 & 08$^{\rm h}$54$^{\rm m}$54$\fs$915677   & $\pm0.048$  & $+$19$\degr$59$\arcmin$34$\farcs$85634  & $\pm0.052$ & 0.38$\pm$0.04 \\
     & S2 & 3.52$\pm$0.08 & 3.06$\pm$0.15 & 08$^{\rm h}$54$^{\rm m}$54$\fs$915709   & $\pm0.097$  & $+$19$\degr$59$\arcmin$34$\farcs$85637 & $\pm0.083$  & 0.52$\pm$0.06 \\
     & S3 & 1.68$\pm$0.03 & 1.81$\pm$0.09 & 08$^{\rm h}$54$^{\rm m}$54$\fs$915737   & $\pm0.138$  & $+$19$\degr$59$\arcmin$34$\farcs$85744  & $\pm0.150$ & 0.29$\pm$0.09 \\
43.2 & S1 & 2.07$\pm$0.07 & 4.09$\pm$0.48 & 08$^{\rm h}$54$^{\rm m}$54$\fs$915677   & $\pm0.010$  & $+$19$\degr$59$\arcmin$34$\farcs$85634  & $\pm0.013$ & 0.31$\pm$0.04 \\
     & S0 & 0.60$\pm$0.09 & 1.54$\pm$0.51 & 08$^{\rm h}$54$^{\rm m}$54$\fs$915677   & $\pm0.028$  & $+$19$\degr$59$\arcmin$34$\farcs$85634  & $\pm0.036$ & 0.35$\pm$0.10 \\
\enddata
\tablecomments{Columns are as follows:
(1) Observing frequency;
(2) component name;
(3) peak brightness;
(4) integrated flux density;
(5) Right ascension;
(6) 1 $\sigma$ of the formal uncertainty in Right Ascension;
(7) Declination;
(8) 1 $\sigma$ of the formal uncertainty in Declination;
(9) component size
}
\end{deluxetable}

\begin{deluxetable}{ccccccccc}
\centering
\tablecolumns{9}
\tabletypesize{\small}
\tablewidth{0pt}
\tablecaption{\label{model} Circular-Gaussian model-fitting results}
\tablehead{
\colhead{Date} & \colhead{Project} & \colhead{Array} & \colhead{Freq.} & \colhead{Component} & \colhead{S$_{\rm tot}$}  & \colhead{R} & \colhead{P.A.} & \colhead{$\theta$}  \\
\colhead {(yyyy-mm-dd)} & \colhead{code} & \colhead{} & \colhead {(GHz)} & \colhead{(mas)} & \colhead {(mJy)}       & \colhead {(mas)}    & \colhead {($\degr$)}        & \colhead {(mas)}      \\
\colhead{(1)}   & \colhead{(2)}  & \colhead{(3)}       & \colhead{(4)}           & \colhead{(5)}         & \colhead{(6)} & \colhead{(7)} & \colhead{(8)} & \colhead{(9)}}
\startdata
2012-03-07 & EP081A  & EVN  & 5.0 &  S  & 28.16$\pm$2.12 & ...           & ...   & 0.65$\pm$0.08  \\
           &         &      &     &  N  & 8.75$\pm$0.79  & 5.70$\pm$0.15 & 39.81 & 0.88$\pm$0.29  \\
2012-03-09 & BD161AS & VLBA & 1.4 & N+S & 84.86$\pm$4.24 & ...           & ...   & 1.98$\pm$0.14  \\
2012-09-14 & BP167B  & VLBA & 2.3 &  S  & 51.29$\pm$0.63 & ...           & ...   & 0.87$\pm$0.09  \\
           &         &      &     &  N  & 17.25$\pm$0.44 & 5.71$\pm$0.06 & 38.66 & 0.65$\pm$0.30  \\
           &         &      & 5.0 &  S  & 34.87$\pm$0.31 &...            & ...   & 0.85$\pm$0.13  \\
           &         &      &     &  N  & 10.94$\pm$0.14 & 5.68$\pm$0.12 & 40.31 & 0.92$\pm$0.23  \\
           &         &      & 8.4 &  S  & 21.52$\pm$0.38 &...            & ...   & 1.10$\pm$0.03  \\
           &         &      &     &  N  &  5.99$\pm$0.43 & 5.72$\pm$0.09 & 39.96 & 1.41$\pm$0.12  \\
2020-11-04 & BR235F  & VLBA & 4.3 &  S  & 46.07$\pm$2.56 & ...           & ...   & 0.72$\pm$0.10  \\
           &         &      &     &  N  & 14.37$\pm$1.36 & 5.73$\pm$0.15 & 39.02 & 1.10$\pm$0.26  \\
           &         &      & 7.6 &  S  & 24.89$\pm$1.32 &...            & ...   & 0.44$\pm$0.10  \\
           &         &      &     &  N  &  4.47$\pm$1.54 & 5.67$\pm$0.33 & 40.69 & 0.94$\pm$0.50  \\
\enddata
\tablecomments{Columns are as follows:
(1) the observation epoch;
(2) Project codes of the observation;
(3) Array name;
(4) Observing frequency;
(5) component name;
(6) integrated flux density;
(7) radial distance from the component S;
(8) position angle with respect to S, measured from north through east;
(9) component size;}
\end{deluxetable}

\end{document}